\documentclass[twocolumn,amsmath]{revtex4-1}
\usepackage[utf8]{inputenc}
\usepackage{graphicx}
\usepackage{bm}
\usepackage{textcomp}

\begin{document}
\title{XUV Spectral Compression by Four-Wave Mixing}

\author{L. Drescher}\email{Lorenz.Drescher@mbi-berlin.de}
\author{V. Shokeen}
\author{T. Witting}
\author{O. Kornilov}
\author{M.J.J. Vrakking}
\author{B. Schütte}\email{Bernd.Schuette@mbi-berlin.de}

\affiliation{Max-Born-Institut für nichtlineare Optik und Kurzzeitspektroskopie, Max-Born-Strasse 2A, 12489 Berlin, Germany}
\date{\today}

\maketitle

\textbf{Extreme-ultraviolet (XUV) sources including high-harmonic generation, free-electron lasers, soft x-ray lasers and laser-driven plasmas are widely used for applications ranging from femtochemistry and attosecond science to coherent diffractive imaging and EUV lithography. The bandwidth of the XUV light emitted by these sources reflects the XUV generation process used. While light from soft-x-ray lasers~\cite{rocca-1999} and XUV FELs~\cite{allaria-2012} typically has a relatively narrow bandwidth, plasma sources and HHG sources driven by few-cycle laser pulses emit broadband XUV pulses~\cite{krausz-2009}. Since these characteristic properties of a given XUV source impose limitations to applications, techniques enabling modification of the XUV bandwidth are highly desirable. Here we demonstrate a concept for efficient spectral compression of a broadband XUV pulse via four-wave mixing (FWM) in the presence of a broadband near-infrared (NIR) pulse in a krypton gas jet, exploiting a phase-matching scheme based on closely-spaced resonances. Our concept provides new possibilities for tailoring the spectral bandwidth of XUV beams.
} 

Since the advent of broadband XUV sources, several methods have been developed to select narrowband XUV radiation, including the use of grating monochromators~\cite{poletto-2009}, multilayer mirrors~\cite{hatayama-2016} and opto-optical modulation~\cite{bengtsson-2017}. However, these techniques come at the cost of losing a large fraction of the XUV flux by filtering photons that are outside of the spectral region of interest. An increased spectral intensity in a narrow bandwidth would be interesting for a range of applications, including coherent diffractive imaging (CDI) for studying solid targets with high spatial resolution~\cite{sandberg-2007}, time-  and angle-resolved photoelectron spectroscopy~\cite{sie-2019} or EUV-lithography~\cite{wagner-2010}.  Here, we demonstrate that FWM processes can be used to transfer a broad incident XUV spectrum into a narrowband region, leading to an increase of the spectral intensity. In the presented experiment, we observe an increase of the spectral intensity in a narrowband region. An estimated bandwidth from calculations suggests the observation corresponds to a 30-fold increase in spectral intensity.

Spectral conversion by FWM in the XUV region near resonances has been extensively studied in the past (see e.g. Refs.~\cite{hilber-1987,dorman-1999,cao-2016,cao-2016a,ding-2016a,harkema-2018,fidler-2019a}). The absorption of one XUV photon (at an energy $\hbar\omega_1$) and that of two additional NIR photons is followed by the emission of one XUV photon (at an energy $\hbar\omega_2$). This process can be well understood by considering the third-order polarization response of the mixing medium. The effective bandwidths of the absorbed and emitted radiation are typically limited to narrow spectral regions, in which phase-matching of the incoming and outgoing waves can be achieved. The phase-matching condition is dictated by the frequency-dependent refractive indices $n(\omega)$:
\begin{align}
\Delta k =& n(\omega_2)\frac{\omega_2}{c} - n(\omega_1)\frac{\omega_1}{c}- 2n(\omega_\textrm{NIR})\frac{\omega_\textrm{NIR}}{c}.\label{eq:phasem}
\end{align}

\begin{figure}[tb]
    \includegraphics[width=.5\textwidth]{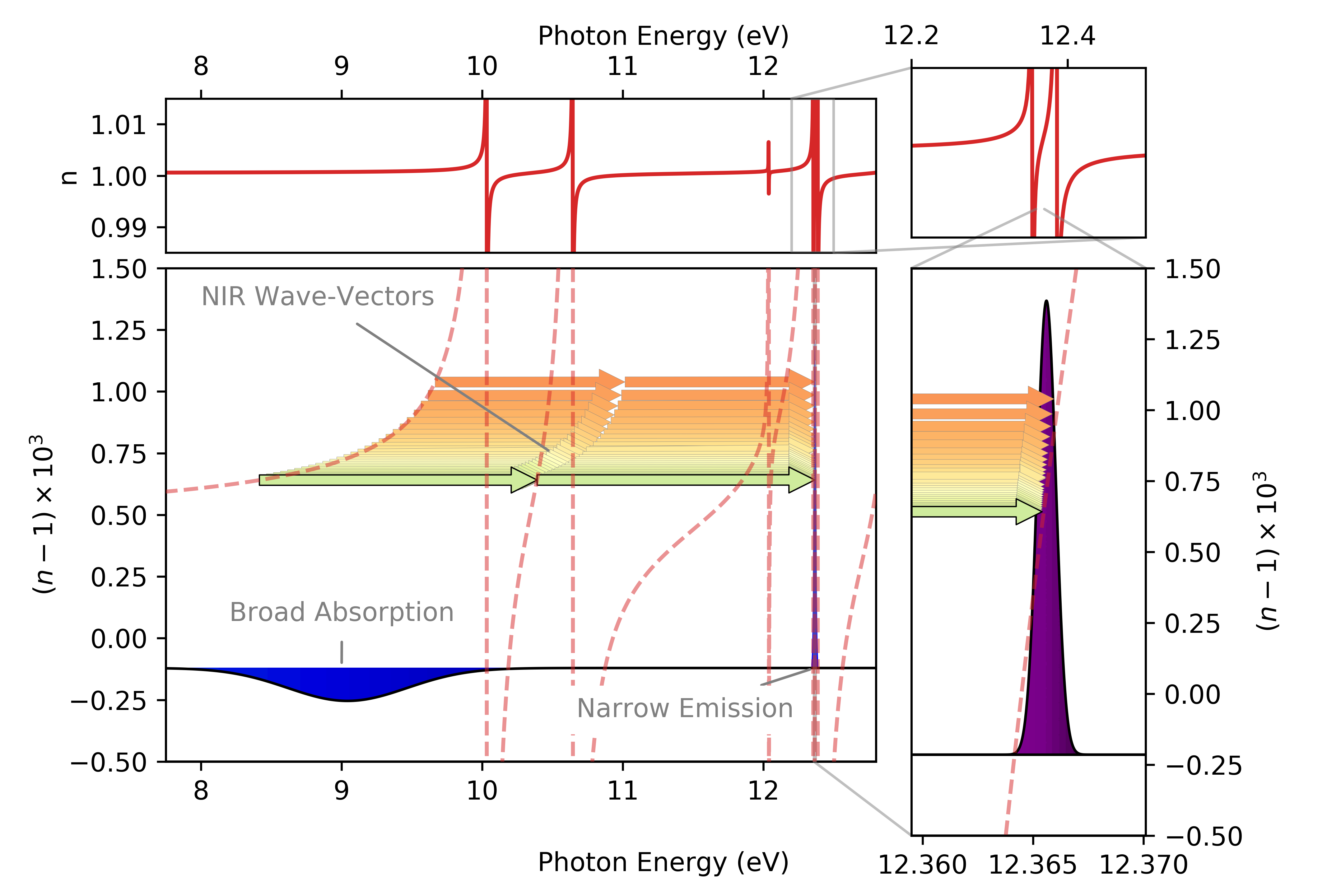}
    \caption{\textbf{XUV Bandwidth compression scheme in Kr:} The extremely different slopes of the energy dependence of the refractive index $n$ at the incident XUV photon energy ($\sim9$\,eV) and at the XUV emission energy ($12.365$\,eV) allow phase-matching of a FWM process involving the absorption of two broadband NIR photons (colored arrows), generating XUV emission within a confined spectral region ($\Delta E \lesssim 2$\,meV) between the 4d and 6s resonances of Kr. The upper panel shows the frequency-dependent refractive index in Kr.\label{fig:scheme}}
\end{figure}

The novel phase-matching concept that we introduce in this paper is presented in Fig.~\ref{fig:scheme}. Bandwidth compression in Kr is achieved by selecting spectral regions in such a manner that the slope of the frequency-dependent refractive index in the absorbing region is small, while it is much steeper in the emitting region. 
The steep slope of the refractive index in the emitting region is defined by its position between two closely spaced resonances. In the vicinity of a single resonance the refractive index typically shows a slow increase approaching the resonance from the lower energy side followed by a quick drop below unity at the resonant frequency and a slow rise back towards $n=1$ at energies above the resonance. In between two neighboring resonances, however, the refractive index crosses unity exhibiting a very steep increase. This steep increase can be exploited for phase-matching incident XUV frequencies over a wide range of photon energies, while producing XUV frequencies within a very narrow photon energy window only.

In Kr, the slope of the refractive index $n$ at the emission photon energy of 12.365\,eV, between the Kr 4d and 6s resonance, is three orders of magnitude larger than the slope of $n$ at $\sim9$\,eV, i.e. below the Kr 5s resonance. The difference in energy matches that of approximately 2 NIR photons. Taking into account the near-constant refractive index in the NIR region ($n_\textrm{NIR}\approx1.0004$~\cite{bideau-mehu-1981}), both regions can be phase-matched in the presence of a few-cycle NIR pulse, $\Delta E\sim1$\,eV, over a broad range of incident XUV photon energies, while producing XUV radiation with a very narrow spectral bandwidth in the emission region ($\Delta E\lesssim$2\,meV). Since this scheme enables absorption of radiation over a broad range of wavelengths but confines the emission to a very narrow range, it leads to a dramatic increase in spectral intensity at the emission wavelength. Moreover, since the refractive index crosses unity steeply away from the resonances that are responsible for the frequency dependence of the refractive index (i.e. where the residual third-order susceptibility is high and absorption by the medium is low), the phase-matching condition can be maintained over a large pressure-length range, leading to an efficient FWM process.

\begin{figure}[tb]
    \includegraphics[width=.5\textwidth]{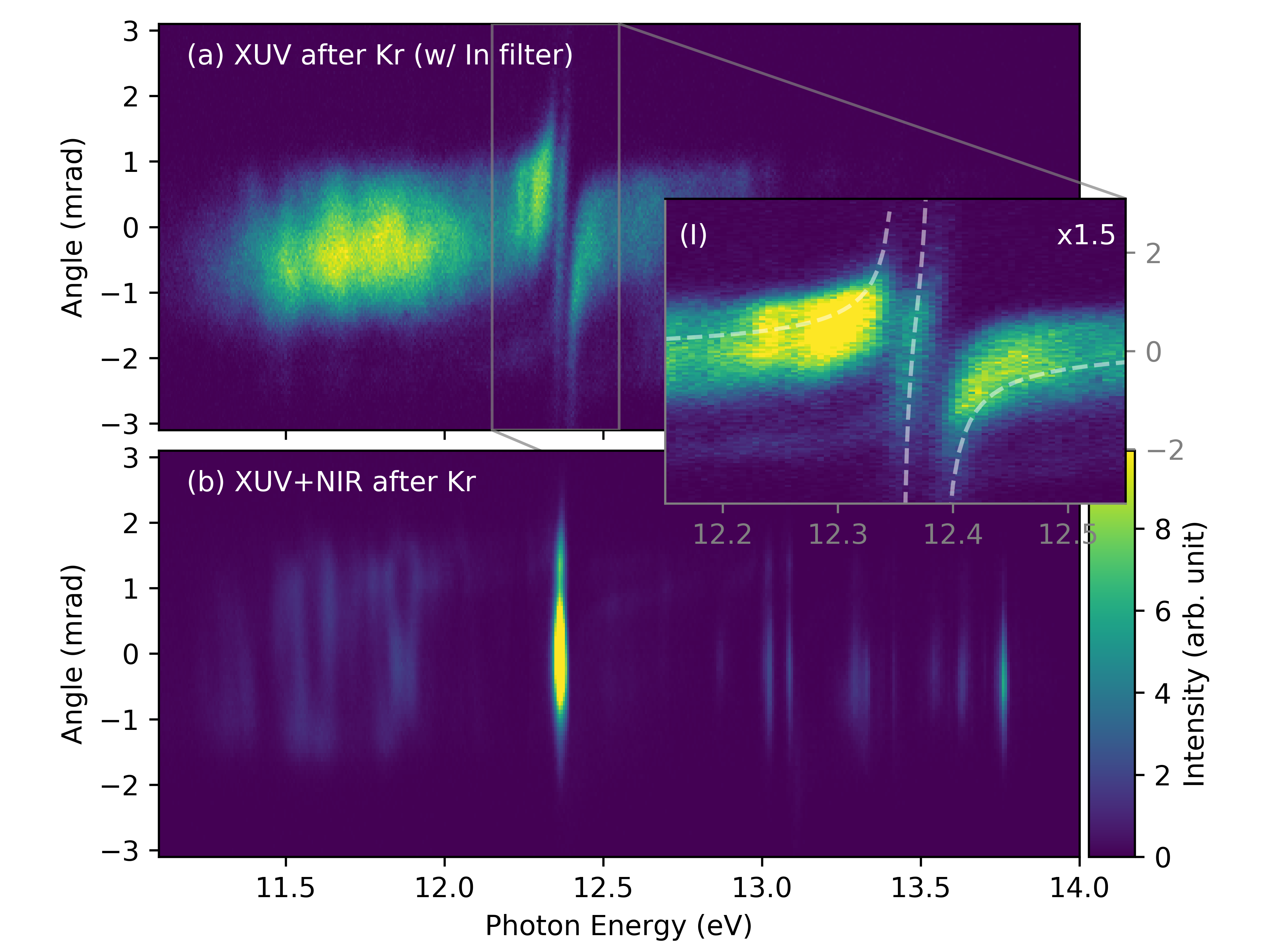}
    \caption{\textbf{Experimental demonstration of XUV compression by means of FWM}: \textbf{a} Angle-resolved XUV spectrum (measured after transmission of an XUV pulse created by HHG through an In filter) after propagation 0.3\,mm below the center of a Kr gas jet ($\approx 3 \times 10^{19}$\,atoms/cm$^3$). The XUV spectra are deflected around Kr resonances because of a wavefront rotation that results from the inhomogeneous transverse density profile~\cite{drescher-2018}.  A narrow undeflected band can be seen between the 4d and 6s resonance (as indicated by the dashed white line in inset \textbf{I}). This narrow feature occurs at photon energies where the XUV refractive index is close to unity. \textbf{b} In the presence of both XUV and NIR laser fields (w/o In filter) a strong emission is observed at the frequencies of this narrow spectral band.\label{fig:exp}}    
\end{figure}

To demonstrate this concept experimentally, we make use of a gas-phase refraction prism, as recently described~\cite{drescher-2018}. Fig.~\ref{fig:exp}(a) shows a measured angle-resolved spectrogram after a broadband XUV pulse obtained by HHG passes 0.3\,mm below the center of a dense Kr gas jet (see Methods). After passing through a region with a gas density gradient, the spectral components are deflected proportionally to the deviation of the frequency-dependent refractive index from unity.
The spectral range shown in Fig.~\ref{fig:exp} is centered around the ($^2P_{3/2}$)4d (12.355\,eV) and ($^2P_{3/2}$)6s (12.385\,eV) resonances~\cite{saloman-2007}, i.e. the expected emission region. The XUV spectral components close to the 4d and 6s resonances are strongly deflected, however, as can be seen from the inset, in between the resonances the refractive index (and hence, the deflection) has a very steep slope and passes through $n=1$ at 12.365\,eV, leading to a narrow feature displaying small deflection angles between the two resonances.

\begin{figure}[tb]
    \includegraphics[width=.5\textwidth]{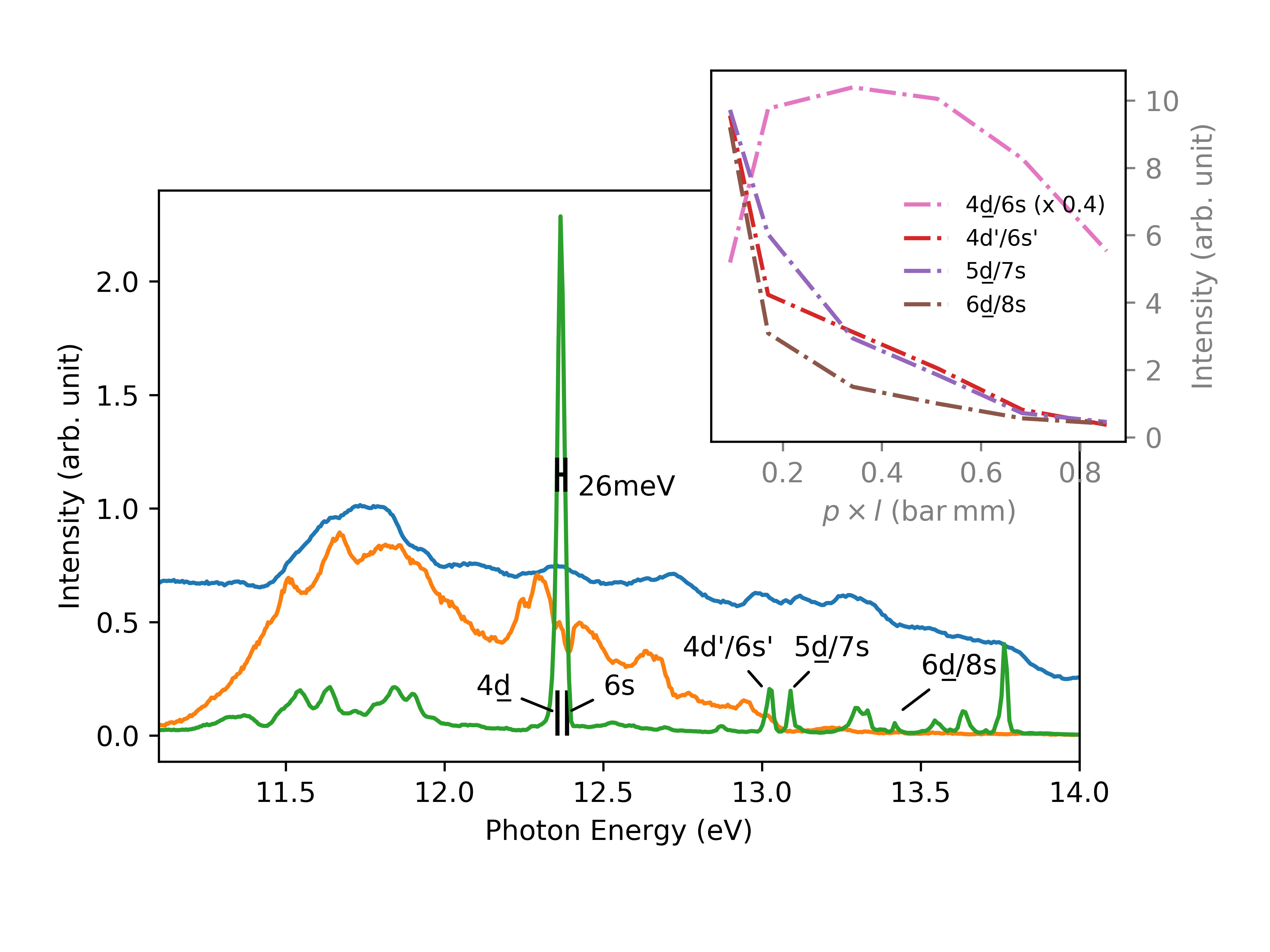}
     \caption{\textbf{Comparison of XUV spectral intensities:} Measured spectra integrated over an angular range between $-3$\,mrad and $+3$\,mrad. The spectral intensity of the narrow feature obtained after co-propagation with the NIR pulse through the gas jet (green) significantly exceeds the spectral intensity measured without gas jet (blue) or in absence of the NIR pulse (orange, measured with In filter). The narrow feature has a spectrometer resolution-limited width of about 26\,meV and a peak position of $12.365 \pm 0.015$\,eV. Additional weak emission features are attributed to FWM near other resonances~\cite{saloman-2007} and show a distinctively different dependence on the gas pressure (inset).\label{fig:exp_b}}
\end{figure}

When a moderately intense ($I\approx 1 \times 10^{12}$~W/cm$^2$) NIR pulse is temporally overlapped with the incident XUV pulse, a strong increase in XUV light intensity is observed (see Fig.~\ref{fig:exp}(b)) at the same photon energy (12.365\,eV). This peak has a measured width of 26\,meV, limited by the spectrometer resolution. Away from this peak, incoming XUV radiation is efficiently absorbed due to NIR-induced coupling of excited states to other states and ionization continua~\cite{schutte-2020}.
Importantly, as can be seen in Fig.~\ref{fig:exp_b}, the measured spectral intensity at the emission feature significantly exceeds that of the incident XUV radiation. This is consistent with the proposed XUV bandwidth compression scheme. 
We note that no signal is observed on the detector at these NIR intensities in the absence of the XUV pulse.

Similar, but much weaker features are observed at lower and higher photon energies. However, their dependence on the gas pressure is distinctively different from that of the peak at $12.365$\,eV: Varying the pressure-length product of the Kr medium  (inset of Fig.~\ref{fig:exp_b}), a pronounced increase in intensity of the feature at $12.365$\,eV is observed, with a maximum occurring at a pressure-length product of about 0.4\,bar\,mm. In contrast, the other observed emission features are rapidly suppressed with increasing pressure. This indicates that the phase-matching condition for bandwidth compression is not fulfilled for these features.

\begin{figure}[tb]
    \includegraphics[width=.5\textwidth]{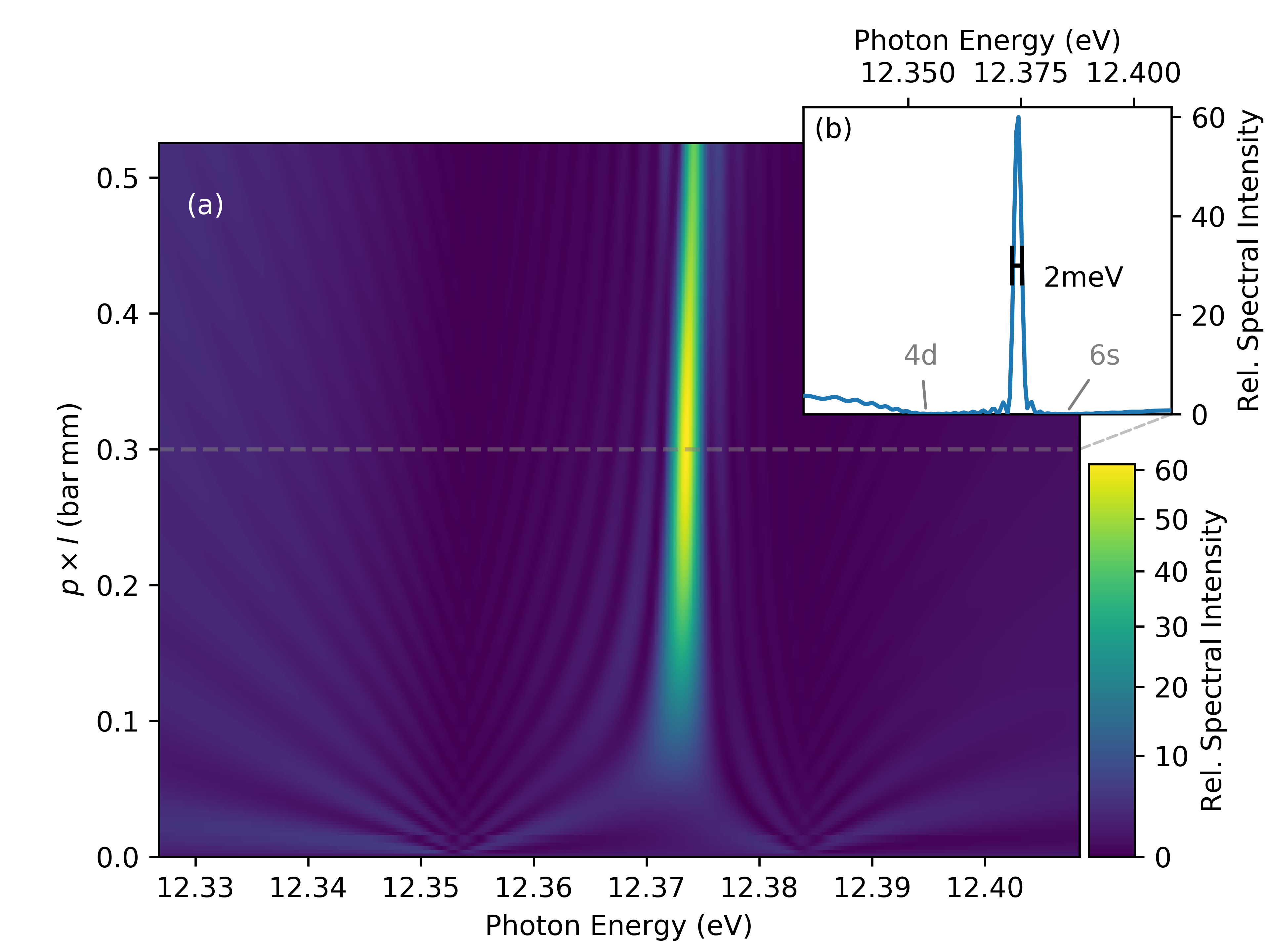}
    \caption{\textbf{Simulated propagation-dependent intensity of the narrowband feature}: (a) Calculated XUV spectra co-propagating with an NIR pulse through a one-dimensional Kr gas using coupled TDSE and Maxwell propagation (see Methods) are plotted for different pressure-length products of the gas medium. While at small propagation depths a broad emission between the two resonances can be seen, the bandwidth quickly narrows upon further propagation. Similar to the experiment, a saturation of the narrow feature is observed at a pressure-length product of about 0.3\,bar\,mm. (b) At its maximum, the narrow-bandwidth emission feature at 12.37\,eV between the strongly absorbing 4d and 6s resonances shows a 60-fold increase of spectral intensity compared to the incident XUV spectrum near 9\,eV and a bandwidth of 2\,meV.}\label{fig:tdse}
\end{figure}

In order to further investigate the underlying mechanism that leads to the spectral compression, we solve the time-dependent Schrödinger equation (TDSE) coupled to Maxwell's wave equation (MWE) for one-dimensional propagation of the two-color XUV and NIR laser fields through a dense Kr gas medium (see Methods).
In Fig.~\ref{fig:tdse}(a), the XUV spectrum after propagation of the two-color laser field as a function of the pressure-length product of the Kr gas is shown. In agreement with the experimental findings, a strong and narrow emission feature between the 4d and 6s resonance is observed. Since only the optically strong resonances of the $J=1$ series of Krypton are included in the TDSE calculation, the feature is slightly offset in energy from the experimental observation and the scheme in Fig.~\ref{fig:scheme}. As can be seen in the inset (Fig.~\ref{fig:tdse}(b)), at a pressure-length product of 0.3\,bar\,mm, the feature has a bandwidth of $\approx$2\,meV (FWHM), which is consistent with the estimate in Fig.~\ref{fig:scheme}. This bandwidth would imply a 30-fold spectral increase in the resolution limited experimental results (Fig.~\ref{fig:exp_b}). When comparing the peak spectral intensity of the emission to the incoming spectral intensity at $\hbar\omega_\textrm{1}\approx9.27$\,eV, a 60-fold increase is found. With an estimated absorption bandwidth that follows from the NIR bandwidth of $\approx0.2$\,eV, this indicates an efficiency of the conversion process of up to $60\%$ (see Methods).
At pressure-length products $>$0.3\,bar\,mm the intensity of the narrowband feature decreases. This decrease is attributed to depletion of the spectral power at $\hbar\omega_\textrm{1}\sim9$\,eV in combination with re-absorption of the generated XUV emission.

The non-resonant nature of the observed process is confirmed by observing the delay dependence of the emission feature: Both, the experiment and the simulation show a fast reduction of the emission intensity outside of the temporal overlap of the XUV and NIR pulses. A perturbative treatment using monochromatic plane waves confirms the suggested phase-matched, non-resonant FWM process: Absorption of XUV radiation around 9\,eV (i.e. below the ($^2P_{3/2}$)5s excited state at $10.032$\,eV)~\cite{saloman-2007}) and of two NIR pulses, aided by the ($^2P_{3/2}$)5p excited states (between $11.3$\,eV and $11.7$\,eV), leads to a strong third-order polarization response between the 4d and 6s excited states (see Supplementary Material). Indeed, exclusion of any of these states in the TDSE quenches the observation of the emission feature.

The presented simulation shows a rapid extinction of other emission features in the vicinity of the 4d and 6s resonances after short propagation lengths due to imperfect phase-matching and absorption. Similarly, at higher photon energy resonances, such as the 5d and 7s resonances, exhibit no build-up of a narrowband emission at high pressure-length products (see Supplementary Material), in line with the experimental results. 
The lines observed in the vicinity of higher-lying resonances are instead attributed to FWM as investigated previously~\cite{hilber-1987,dorman-1999,cao-2016,cao-2016a,ding-2016a,harkema-2018,fidler-2019a}.

In conclusion, we have presented an XUV spectral compression scheme, in which broadband radiation from an attosecond pulse train is converted into XUV radiation with narrow bandwidth. The residual third-order susceptibility at XUV frequencies between two electronic resonances in Kr leads to a strong polarization response and a narrow-bandwidth emission feature with high spectral flux. The presented theoretical models show that a phase-matching condition enables this process to be efficient at high pressure-length products, where absorption is nevertheless low. In view of its dependence on the nature of refractive index changes around resonances, the underlying phase-matching mechanism is expected to be applicable whenever such features can be found, i.e. in other atomic and molecular gases at different frequencies. The use of ion plasma could allow the extension of this scheme into the high energy XUV or even soft-x-ray ranges~\cite{chu-2018}. 

Although in this work we have used the presented mechanism to generate a narrow-bandwidth emission feature starting from an available broad bandwidth XUV source, the non-resonant FWM can be reversed: Using comparably narrow-bandwidth XUV pulses the presented scheme could be used to generate XUV pulses with a larger bandwidth. Our method thereby also opens a way to compress XUV pulses to shorter durations, similar to hollow-core fiber compression in the visible regime. This could enable development of ultrahigh repetition rate sources of broadband XUV pulses based on commercially available Ti:sapphire oscillators by utilizing below-threshold harmonic generation~\cite{chini-2014b} or other narrow-bandwidth XUV sources, such as seeded FELs or soft x-ray lasers.

\appendix*
\section*{Methods}
\subsection{Experiment}
Short NIR pulses of 4.5\,fs pulse duration (FWHM), centered around 750\,nm wavelength, are used to generate high harmonics in Xe, using a beamline that was previously described in detail~\cite{drescher-2018,drescher-2019,galbraith-2017b}. After recombination with a delay-controlled replica of the NIR beam, both beams are refocused into a pulsed dense Kr gas jet, which propagates horizontally and orthogonally to the laser beam propagation direction. The gas jet can be shifted in the vertical direction to achieve the gas-prism configuration as recently introduced~\cite{drescher-2018}. After interaction with the gas jet, the XUV light is spectrally dispersed by a flat-field grating onto a detector comprised of a multichannel plate and phosphor screen. The signal from the screen is read out by a digital CCD camera. Due to geometrical constraints, the lowest photon energy detectable by the spectrometer setup is 11\,eV. To ensure that the measurements of the XUV spectrum in the absence of the NIR pulse is not influenced by the presence of the residual NIR radiation used in the HHG source, a thin Indium filter is introduced after HHG to block residual NIR photons, as well as XUV light below 11\,eV and above 16\,eV. The XUV spectrum after propagation through the gas jet in the absence of the NIR pulse is qualitatively the same with and without the In filter. To quantitatively compare the measured spectra, all measurements with the In filter have been scaled by matching the XUV spectral intensity in the absence of the gas jet in the region of interest measured with and without In filter. In the experiment, the pressure-length product could be varied by changing the backing pressure of the jet nozzle from 1.1\,bar to 10\,bar. The experimental pressure-length products are estimated by assuming a peak pressure in the gas jet of a 20\% of the backing pressure and a parabolic jet with diameter of 1\,mm. The experiments are performed at a vertical distance from the center of the jet of 0.3\,mm, leading to a further reduction of pressure by a factor of 1.56 due to the parabolic jet expansion. This leads to pressure-length products in the interaction zone from 0.1\,bar\,mm to 0.9\,bar\,mm.

\subsection{TDSE Calculations}
In the TDSE calculations, the time-dependent dipole $d(t)$ resulting from the interaction of the two-color laser field with ground state Kr atoms is calculated using a finite basis set of states from the $J=1$ series and up to an effective quantum number $n^\ast\leq10$. The field-free energies are obtained from Ref.~\cite{nist}, while the transition dipole moment matrix elements are calculated using standard angular momentum algebra and the method for evaluation of the radial matrix elements described in Ref.~\cite{klarsfeld-1988}. After a windowed Fourier transform to account for finite spectral resolution, the dipole spectrum $\tilde d(\omega)$ is used to calculate the modification of the two-color laser field $\tilde S(\omega)$ in the $z$-direction according to~\cite{wu-2016}:
\begin{equation}
d\tilde S(\omega) = \frac{i\rho\omega}{\epsilon_0c}\tilde d(\omega) dz\label{eq:prop}
\end{equation}
where $\rho$ is the atomic density. The TDSE is integrated over 17\,ps, whereas the integration step during the propagation described by Eq.~\ref{eq:prop} is typically $10^6\,\textrm{a}_0$ ($\approx53$\,\textmu m) at a gas pressure of 1\,mbar (assuming a temperature of $300$\,K) for different integration lengths. At the start of the simulation the NIR pulse is assumed to be a 15\,fs FWHM sine-squared pulse centered at 800\,nm with a peak intensity of $9\times10^{11}$\,W/cm$^2$ ($F_\textrm{NIR} = 0.005$\,a.u.), whereas the XUV pulse is assumed to be a Gaussian pulse with a FWHM bandwidth of 0.167\,a.u. (corresponding to a 400\,as FWHM pulse duration), centered at the 7th harmonic of the NIR (10.85\,eV) and with a peak intensity of $3.51\times10^{10}$\,W/cm$^2$. As in the experimental observation, we note that the result of the simulation is robust to variations in the parameters defining the XUV and NIR pulses, as well as to the inclusion of single-photon ionization by the NIR laser (using ionization rates calculated for Ar Rydberg states as described in Ref.~\cite{muller-1999}).

To estimate the efficiency of the spectral compression, the peak spectral intensity of the XUV narrow-bandwidth feature and its bandwidth are compared to the average spectral intensity in the absorption region ($\approx$9.27\,eV) and the available two-photon NIR bandwidth. The $15$\,fs NIR-pulse with central frequency of $1.55$\,eV corresponds to a bandwidth of approximately $0.14$\,eV. Although a sine-squared envelope was used in the calculations, we estimated the two-photon bandwidth by convolution of two Gaussian distributions with this bandwidth. The estimated two-photon bandwidth is then $\Delta E_\textrm{2NIR}=\sqrt{2}\cdot0.14$\,eV). The peak spectral intensity of the emission after a pressure-length product of 0.3\,bar\,mm shows a 60-fold increase of the spectral intensity compared to the incoming spectral intensity at $\hbar\omega_\textrm{1}\approx9.27$\,eV, while the spectral width is reduced by a factor 100, from $\approx\sqrt{2}\cdot0.14$\,eV to $\approx2$\,meV. This indicates that up to $\sim60\%$ of photons in the absorption region are compressed into the emitted narrow-bandwidth feature.

\section*{Acknowledgments}
The authors would like to thank A. A. Ünal for his support with the laser system and M. Ivanov, M. Richter, F. Morales, A. Husakou and S. Patchkovskii for helpful discussion.

\bibliography{main}
\end{document}


\title{XUV Spectral Compression by Four-Wave-Mixing (Supplementary Material)}

\maketitle

\section{Perturbative Treatment of Third-Order Polarization Response}

Perturbation theory is used to calculate the third-order susceptibility for monochromatic waves in the atomic unit system evaluating the sum over excited states~\cite{boyd-2013,miles-1973}:
\begin{equation}
\chi^{(3)}(\omega)=\frac{N}{\epsilon_0} \sum_{abc}{d_{ga}d_{ab}d_{bc}d_{cg}}A^{(3)}_{abc}
\end{equation}
where $d_{nm}$ are the transition dipole moments, $N$ is the number density, $\epsilon_0$ the dielectric constant and
\begin{equation}
A^{(3)}_{abc} \approx \mathcal{P}_{ijk}\frac{1}{(\tilde\omega_{ag}-\omega_i-\omega_j-\omega_k)(\tilde\omega_{bg}-\omega_j-\omega_k)(\tilde\omega_{cg}-\omega_k)}\label{eq:freqfact}
\end{equation}
with $\tilde\omega_{mn} = \omega_{mn}-i\Gamma$ the field-free state energies with an added imaginary energy to account for their natural linewidth, and $\mathcal{P}$ marking the sum over the permutations of the incoming waves i,j and k corresponding to $\omega_\textrm{XUV}$ and (twice) $\omega_\textrm{NIR}$. 
The field-free state energies for a finite basis set states  belonging to the $J=1$ series up to an effective quantum number $n^\ast\leq10$ are obtained from Ref.~\cite{nist}, while the transition dipole moment matrix elements are calculated using standard angular momentum algebra and the method for evaluation the radial matrix elements described in Ref.~\cite{klarsfeld-1988}. 
Since we assume the Kr atoms to be in the ground state, the counter-rotating terms are neglected in the evaluation of Eq.~\ref{eq:freqfact}. Due to conservation of energy $\omega_\textrm{XUV,2} = \omega_\textrm{XUV,1}+2\omega_\textrm{NIR}$.
Likewise for the phase-matching condition:
\begin{align}
\begin{split}
\Delta k =& n(\omega_\textrm{XUV,2})\frac{\omega_\textrm{XUV,2}}{c} \\ &- n(\omega_\textrm{XUV,1})\frac{\omega_\textrm{XUV,1}}{c}- 2n(\omega_\textrm{NIR})\frac{\omega_\textrm{NIR}}{c}.\label{eq:phasem}
\end{split}
\end{align}
First-order perturbation theory is used to calculate the linear susceptibility~\cite{boyd-2013}:
\begin{equation}
\chi^{(1)}(\omega) = \frac{N}{\epsilon_0}\sum_a \frac{d_{ag}d_{ga}}{\omega_{ag}-\omega},\label{eq:lin_chi}
\end{equation}
and the refractive index $n=\sqrt{1+\chi^{(1)}(\omega)}\approx 1+\frac{1}{2}\chi^{(1)}(\omega)$.

\begin{figure}[tbp]
    \includegraphics[width=\textwidth]{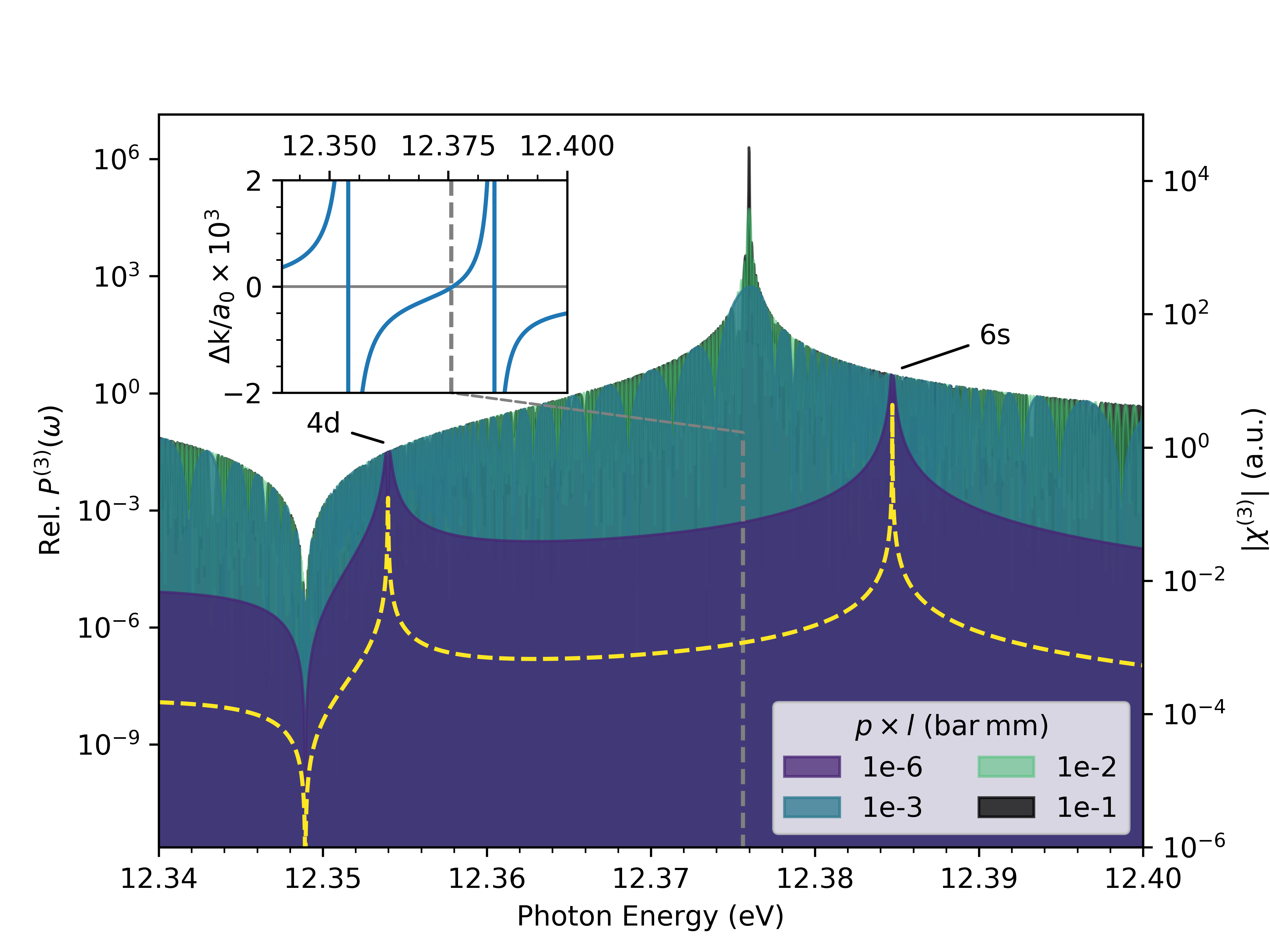}
    \caption{\textbf{Polarization Response and Phase-Matching:} The third-order susceptibility ($\chi^{(3)}(\omega_\textrm{2})$), dashed yellow line) of Kr in the vicinity of the 4d and 6s resonances and the zero-crossing of the wave-vector mismatch described by Eq.~\ref{eq:phasem} (inset) allow for an effective FWM process. For large pressure-length products ($p\cdot l > 10^{-2}$\,bar\,mm), a narrow-bandwidth emission feature emerges in the relative third-order polarization response.}\label{fig:perturb}
\end{figure}

The polarization response of the FWM process is calculated, which for monochromatic plane-waves can be described as~\cite{miles-1973}:
\begin{equation}
\frac{P^{(3)}(\omega_\textrm{2})}{P_\textrm{1}P_\textrm{NIR}^2} \propto l^2[\chi^{(3)}(\omega_\textrm{2})]^2\operatorname{sinc}\left(\frac{\Delta k(\omega_\textrm{2})l}{2}\right)^2,\label{eq:relp}
\end{equation}
where $P_\textrm{1},P_\textrm{NIR}$ are the respective monochromatic intensities, $l$ the interaction length and $\chi^{(3)}(\omega_\textrm{2})$ and $\Delta k(\omega_\textrm{2})$ are the frequency- and (approximately linearly) pressure-dependent third-order susceptibility and wave-vector mismatch (see Eq.~\ref{eq:phasem}). The polarization response in the vicinity of the 4d and 6s resonances is shown for different pressure-length products in Fig.~\ref{fig:perturb} for a fixed NIR photon energy of 1.55\,eV. For low pressure-length products ($p\cdot l\leq 10^{-6}$bar\,mm) $P^{(3)}(\omega_\textrm{2})$ follows the third-order susceptibility (dashed yellow curve). For high pressure-length products ($p\cdot l\geq 10^{-1}$bar\,mm), however, we observe the build-up of a narrowband emission at a spectral position that is determined by the wave-vector mismatch (inset in Fig.~\ref{fig:perturb}). The omission of Kr states not belonging to the $J=1$ series leads to a small energy shift of the narrow phase-matched region, when compared to the experimental results.

\section{Behaviour of Four-Wave-Mixing near the 5d and 7s resonances}
\begin{figure}[tbhp]
    \includegraphics[width=\textwidth]{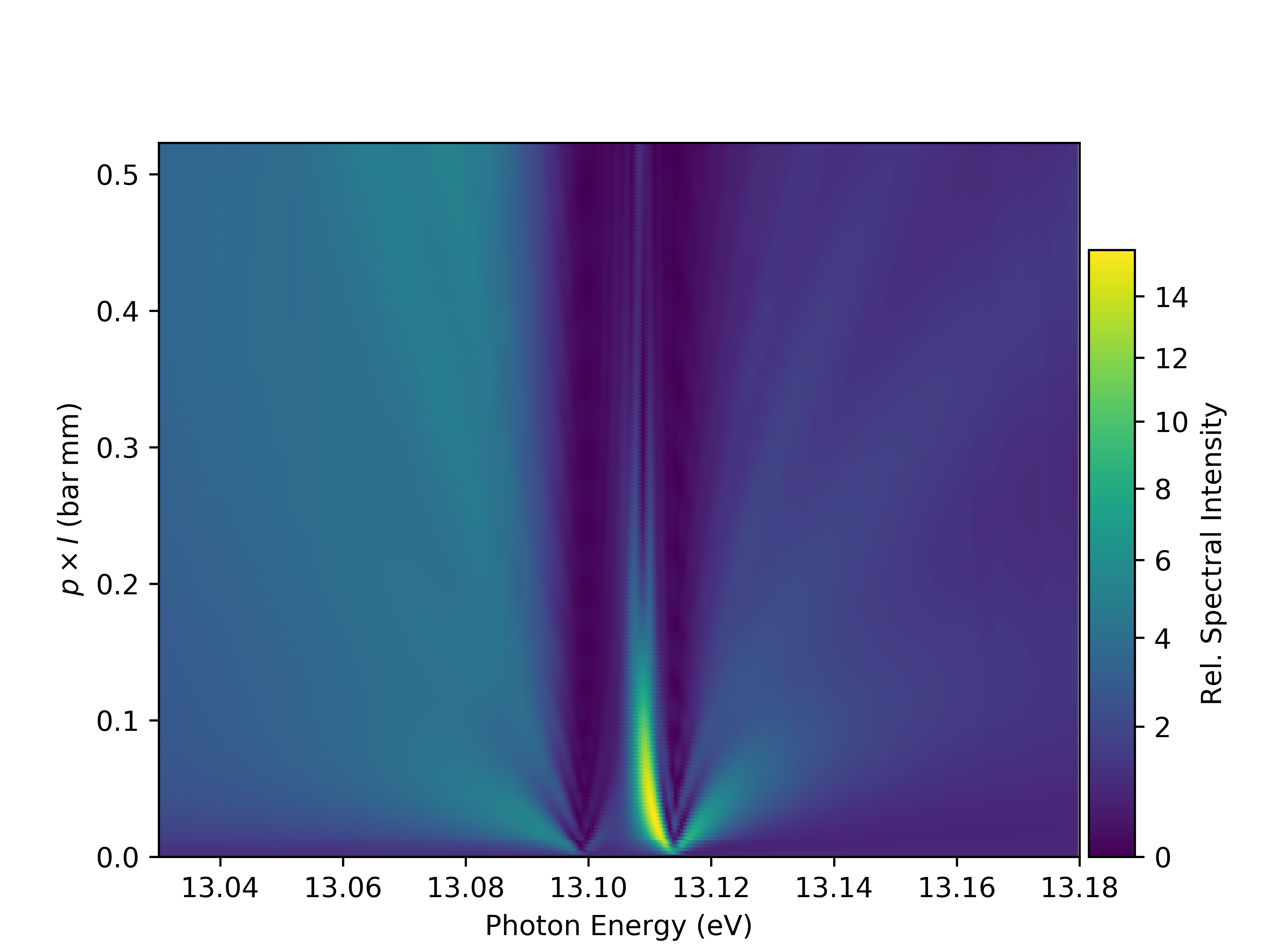}
    \caption{\textbf{Simulation of the FWM signal near the Kr 5d and 7s resonances}: XUV spectra after co-propagation with an NIR pulse through a one-dimensional Kr gas calculated by solving the coupled TDSE and Maxwell propagation equations. In line with the experimental observation, a narrowband emission feature is only observed in this spectral region for a low pressure-length product $p\cdot l<0.1$\,bar\,mm. The efficient formation of a narrowband emission feature is therefore not observed.}\label{fig:tdse_more}
\end{figure}

In line with the experimental observation, the TDSE and MWE model shows a different behaviour of the FWM near e.g. the 5d and 7s resonant energies, when compared to the narrow emission feature between the 4d and 6s. In Fig.~\ref{fig:tdse_more} we show the evolution of the XUV spectrum during the propagation through the jet for the spectral region around the 5d and 7s resonances. While the formation of an emission feature between the 5d and 7s resonances can be seen for small propagation lengths (i.e. around $13.11$\,eV at a pressure-length product of below $0.1$\,bar\,mm), this emission feature is not observed at larger pressure-length products. The behaviour is closer to the observations reported in previous studies of XUV FWM (see e.g. Refs~\cite{cao-2016,cao-2016a}). The different behaviour can be explained by the phase-matching condition. While the narrow bandwidth feature between the 4d and 6s resonances can be phase-matched to a broad region below the Kr 5s resonance by two NIR photons, in the case of the 5d and 7s region the same NIR photons will energetically map onto a region in the vicinity the 5s resonance, which leads to a phase-matching condition that is critically dependent on the pressure-length products. This, however, implies that a similar bandwidth compression can be achieved at an energy between the 5d and 7s resonant energies, if an appropriate beam in the visible spectral range is used, so that the emission region is energetically matched to an absorption region that is further below the Kr 5s resonance.

We note that the described scheme is conceptually similar to electromagnetically induced transparency~\cite{harris-1990}, in that it describes an enhancement of a non-linear effect due to the existance of a transparent region and a refractive index close to unity. However, it is different in the sense that the scheme described here exploits this configuration in a spectral range where this transparancy is an inherent property of the gas medium, forgoing the need of a resonantly tuned control laser field and allowing broadband matching of the absorbed frequency ranges that allows for an efficient compression of broadband sources into narrowband emission.

\bibliography{support}